\input harvmac

\Title{}{Neutrinos With Seesaw Masses and Suppressed Interactions}

\centerline{\bf Sheldon Lee Glashow}\bigskip

\centerline{Department of Physics}
\centerline{Boston University}
\centerline{Boston, MA 02215}
\vskip .4in

Mixing between light and heavy neutrino states has been proposed as an
explanation (or partial explanation) for  the 3-sigma NuTeV anomaly and the
2-sigma departure of the $Z^0$ invisible width from its expected value.
I assume herein that neutrino masses and mixings result from the
 conventional seesaw mechanism involving six
chiral neutrino states, the first
three being  members of weak doublets, the others
weak singlets. A finely-tuned choice 
 of both the (bare) Majorana
masses and the (Higgs-induced) Dirac masses 
 can fit solar and atmospheric  neutrino data and also
 result in significant (but necessarily flavor-dependent) mixing
between singlet and doublet states such as would yield detectable suppression
of  light neutrino interaction amplitudes. The possibility for this kind
of suppressive mixing is constrained by the observed upper limit
on radiative muon decay.

\Date{01/03}

The three-sigma
 NuTeV anomaly~\ref\rnutev{K.S. McFarland {\it et al.,} {\tt
 hep-ex/0205081}\semi G.P. Zeller {\it et al.,} {\tt hep-ex/0207037.}},
as well as the two-sigma anomaly~\ref\rz{The LEP Collaborations, 
the LEP Electroweak Working Group, and the SLD Heavy Flavor Group, {\tt
hep-ex/0212036.}}
 in the invisible width of the $Z$,
may require new physics for their resolution. Several  proposed
 explanations invoke  significant  mixing of active neutrinos with 
 heavy singlet states~\ref\numix{{\it E.g.,}
 J. Bernard\'eu {\it et al.,} Phys. Lett B187(1987)303\semi
A. de Gouv\^ea {\it et al.,} Nucl. Phys. B623(2002)395, {\tt
hep-ph/0107156}\semi K.S. Babu and J.C. Pati, {\tt hep-ph/0203029}.}.
The effect of this mixing is  to reduce the 
couplings of $W$ and $Z$ bosons to the light neutrinos: 
for  $W$  by the factors $1-\epsilon_\ell/2$, and for 
 $Z$ by the factors $1-\epsilon_\ell$ (where $\ell=
e,\mu,\tau$ is a flavor index).

Davidson {\it et al.}~\ref\sd{S. Davidson {\it et al.,} JHEP 0202(2002)037,
 {\tt hep-ph/0112302}.} find  that neutrino mixing can explain both
 anomalies, but only at the unacceptable cost  of introducing 
additional  departures
 from standard-model predictions. Recently,
 Takeuchi {\it et al.}~\ref\rtak{T. Takeuchi,
 {\tt hep-ph/0209109}; T. Takeuchi {\it et al.,} 2002
International Workshop at Nagoya, 
{\tt www.eken.phys.nagoya-u.ac.jp/SCGT02/.}}~\ref\rloi{ W. Loinaz, N. Okamura,
T. Takeuchi, and L.C.R. Wijewardhana,
{\it et al.,} {\tt hep-th/0210193}.}  
 revived the notion of 
suppressive neutrino mixing in a scheme  also involving an
 unconventionally heavy Higgs boson.
To carry out  their
analyses,  they make the {\it ad hoc\/} hypothesis
of flavor-independent ($\epsilon_\ell=\epsilon$) mixing between  light   and  
heavy neutrino states. Their fit requires
 $\epsilon=0.003\pm 0.001$, about which they remark
that ``a naive
 seesaw mechanism does not permit such a large mixing angle,'' but that
``the required pattern of mixings and masses can be arranged when there
 exists intergenerational mixing''~\rloi.

This work asks whether neutrino mixing of the sort envisaged  by
Takeuchi {\it et al.} can result from a  minimal seesaw mechanism: one
involving exactly
three weak-doublet neutrino states and  exactly three singlet states, 
with Majorana masses for the singlets, Higgs-induced  mass terms  linking  
singlets and doublets, and no weak-triplet Majorana masses of the doublet
states~\ref\rseesaw{S.L. Glashow, in {\it Quarks and Leptons, Carg\`ese
1979,} eds. M. L\'evy, {\it et al.,} (Plenum 1980 New York), p. 707;
M. Gell-Mann, P. Ramond, and R. Slansky, in {\it Supergravity,} eds.
D. Freedman {\it et al.,} (North Holland 1980 Amsterdam); Y. Yanagida in
{\it Proc. Workshop on Unified Theories \&c.,} eds. O. Sawada and A. Sugamoto 
(Tsukuba, 1979); R.N. Mohapatra and G. Senjanovi\'c, Phys. Rev. Lett.
44(1980)912.}.

I find that the neutrino flavor eigenstates,
in an appropriately fine-tuned  model of this kind,  {\it can\/}
mix significantly  with (inaccessible) heavy mass eigenstates, but
this  suppressive  mixing
{\it cannot\/}  be flavor independent. Furthermore, 
I show how the current limit on the
 unobserved decay mode $\mu\rightarrow e+\gamma$ limits  the
nature and magnitude of the permitted flavor-dependent mixings.

 The
most general   $6\times 6$ seesaw neutrino mass matrix is
\eqn\eone{  {\cal M}\equiv \pmatrix{\bf 0&\bf n\cr \bf \tilde n&\bf M\cr} }
where ${\bf n}$  is an arbitrary $3\times 3$ matrix, ${\bf \tilde n}$  is its
transpose, and ${\bf M}$ is an arbitrary $3\times 3$ symmetric matrix. 
I shall assume there to be  a basis in which  these matrices take 
(or nearly  take) the
following special forms:
\eqn\etwo{{\bf n}= \pmatrix{0 &0&0\cr 0&n'&0\cr 0&0&n\cr}\,, \qquad{\rm
  and}\qquad {\bf M}= \pmatrix{\Delta & 0& M\cr
0&M'&0\cr M&0&0\cr}\,, }
where $M,\, M' \gg n,\,n' \gg \Delta>0$. This finely-tuned
neutrino mass matrix is  essentially the only one that can
yield  significant suppressive mixing  in the minimal seesaw scheme.
(The matrix ${\bf n}$ may always be chosen to be diagonal. 
However, the requirements that
$n_{11}$   and certain elements of ${\bf M}$ vanish can be relaxed, but only
to the extent that our conclusions are unaffected.)
   
The approximate eigenvalues of the 
real symmetric matrix $\cal M$ (the neutrino masses) are easily seen to be
$\pm M$ and $M'$ (for the heavy and predominantly weak singlet states) and
$0$, $n^{\prime 2}/M'$ and $n^2\Delta/M^2$ (for the light and predominantly
weak singlet states). Both $M$ and $M'$ must probably exceed
$M_Z$ because no heavy neutrino
state has been detected. 
 I have not as yet specified the basis states of $\cal M$  
in terms of flavor eigenstates, other than by  identifying the first three
components as (active) doublet states and the last three components as
(sterile) singlet states.   
The three light mass eigenstates of $\cal M$  are easily found. 
Properly normalized, they are:

$\bullet\quad \nu_a= (1,\,0,\,0;\;0,\,0,\,0)$: This neutrino state is 
purely  doublet  and massless ($m_a=0$).  

$\bullet\quad \nu_b= (0,\,\cos{\chi'},\,0;\;0,\,\sin{\chi'},\,0)$:  This state,
with mass $m_b\simeq n^{'2}/M'$, mixes negligibly with the singlet 
 states.  (From the  plausible  constraints,
 $m_b<10$~eV and $M'>100$~GeV, I find $n'<1$~MeV and
 $\chi' \simeq n'/M'<10^{-5}$.) Little error is incurred by setting
  $\nu_b\simeq  (0,\,1,\,0;\; 0,\,0,\,0)$.

$\bullet \quad \nu_c\simeq  (0,\,0,\cos{\chi};\;0,\,0,\,\sin{\chi})$: 
This is the only light neutrino eigenstate
which may mix significantly  with singlet  neutrinos.
Its mass is $m_c\simeq n^2\Delta/M^2$, whilst its mixing
parameter is $\chi\simeq n/M$. (For example, we might set $M=1$~TeV,
$n=100$~GeV and $\Delta=10$~eV to obtain
 a neutrino with  a mass
of 0.1~eV, which would then suffer 10\% 
suppressive mixing with singlet  
states.)  We define the parameter $\epsilon$
in terms of which the suppression parameters $\epsilon_\ell$  
corresponding to  the various flavor eigenstates  will be expressed:
\eqn\eps{\epsilon\equiv \sin^2{\chi}\simeq (n/M)^2.}

The basis  in which the light mass eigenstates $\nu_{a,b,c}$ are exhibited
above is one from which inter-family mixing is  removed.  The {\it flavor\/}
eigenstates are 6-component vectors spanning the space of doublet neutrinos.
They are orthonormal vectors whose last three components vanish. 
These states, for
$\epsilon\ne 0$, involve a possibly significant admixture of inaccessible
heavy neutrino states. The `reduced'  flavor
eigenstates,  denoted by ${\hat\nu}_{e,\mu,\tau}$ (the hats indicating
 3-component vectors),  are given
 by a unitary matrix---the usual analog to the Kobayashi-Maskawa
  matrix---acting on ${\hat \nu}_{a,b,c}$, the projections of $\nu_{a,b,c}$
  on the space of doublet states. In other words, $\hat{\nu}_{a,b,c}$ are
  simply the states $\nu_{a,b,c}$ with their last  three components deleted.
In the following equation, the reduced flavor eigenstates are expressed in
terms of a permutation of the states ${\hat\nu}_{a,b,c}$:   
\eqn\km{\pmatrix{{\hat\nu}_e\cr {\hat\nu}_\mu\cr {\hat\nu}_\tau\cr}=
\pmatrix{c_2c_3&c_2s_3&s_2e^{-i\delta}\cr
-c_1s_3-s_1s_2c_3e^{i\delta}&c_1c_3-s_1s_2s_3e^{i\delta}&s_1c_2\cr
+s_1s_3-c_1s_2c_3e^{i\delta}&-s_1c_3-c_1s_2s_3e^{i\delta}&c_1c_2\cr}
\pmatrix{{\hat\nu}_{a'}\cr{\hat\nu}_{b'}\cr 
{\hat\nu}_{c'}\cr}\,,}
where\footnote{${}^\dagger$}{For the angles usually designated by
$\theta_{23},\,\theta_{13},\,\theta_{12}$, we use the simpler names  
 $\theta_1,\,\theta_2,\,\theta_3$, resp.}~$s_i,\,c_i\equiv 
\sin{\theta_i},\,\cos{\theta_i}$, and
$(a',b',c')$ is a permutation of $(a,b,c)$.
 Because $|{\hat\nu}_c|^2=\cos^2{\chi}<1$,
 the reduced flavor eigenstates,
  ${\hat\nu}_e$, ${\hat\nu}_\mu$ and  ${\hat\nu}_\tau$, are neither 
 normalized nor are they orthogonal to one another.

Three phenomenologically distinct permutations  are compatible
with the
observed mass hierarchy  of solar and atmospheric neutrino oscillations.
(The remaining permutations, with $a$ and $b$ interchanged, 
 are accounted for by the variation of $\theta_3$ from $0$ to
$\pi$.)
\medskip

\item{I.} $(a',b',c')=(a,b,c)$: \   where
the parameters in \etwo\ are chosen so  that  $m_c^2\gg m_b^2$.  
For this case, using the above expressions for $\nu_{a,b,c}$ and Eq.\km,
I find the following  suppression factors applicable to the several
flavor eigenstates:
\def\ee{\epsilon}
\eqn\esupe{\ee_e=\ee s_2^2,\qquad\ee_\mu=\ee s_1^2c_2^2,\qquad
\ee_\tau=\ee c_1^2c_2^2.}

\item{II.} $(a',b',c')=(a,c,b)$: \   where
the parameters are chosen so that  $m_b^2\gg m_c^2$.  
For this case, I may neglect the small subdominant angle $\theta_2$
in the expressions for these suppression factors:
\eqn\esupm{\ee_e\simeq\ee s_3^2,\qquad\ee_\mu\simeq \ee c_1^2c_3^2,\qquad
\ee_\tau\simeq \ee s_1^2c_3^2.}

\item{III.} $(a',b',c'=b,c,a)$: \  where the parameters are chosen 
so that $m_c^2\gg 
\vert m_c^2-m_b^2|$. 
For this instance of an inverted mass hierarchy,
the  suppression factors are identical to those of Case II.
(In all three  Cases,  $\epsilon_e+\epsilon_\mu+\epsilon_\tau=1$.) 
\medskip

The departures of the reduced flavor eigenstates from orthonormality,
\eqn\edot{ \big\vert\hat{\nu}^\dagger_\ell\cdot 
{\hat\nu}_{\ell'}\big\vert^2=\ee_\ell\,\ee_{\ell'},
\qquad{\rm where} \ \ \ell,\,\ell' = e,\,\mu,\,\tau, }
 have
 phenomenological consequences other than their relevance to the analyses
of Takeuchi {\it et al.}~\rtak\rloi.
In the seesaw models we have considered, the simple one-loop diagrams
involving the emission and absorbtion of a $W$ boson lead to a nonvanishing
contribution to radiative decays such as  $\mu\rightarrow e +\gamma$. 
This occurs because both the heavy and light components of the neutrino flavor
eigenstates are involved in the virtual process.
It follows from~Eq.\edot\ 
and Ref.\ref\rgf{G. Feinberg, Phys. Rev. 110(1958)1482.}\ 
that the branching ratio $B$ for the radiative decay mode
$\mu\rightarrow e+\gamma$ is:
\eqn\eB{B= {\alpha\over 24\pi}\, \ee_\mu\ee_e\,N(M),}
where the  monotone function  $N(M)$ equals  1/2 at $M=M_Z$,  
increasing to 1 as $M\rightarrow \infty$. This result is to be compared with
the experimental upper bound ~\ref\rbrooks{M.L. Brooks,
  Phys. Rev. Lett. 83(1999)521.}:
\eqn\eex{B<1.2\times 10^{-11}.}
Eqs.\eB\ and \eex\ yield  the inequalities:
\eqn\econst{10^{-6}>\cases{\ee^2\,s_1^2\sin^2{2\theta_2}, & for Case I;\cr  
                           \ee^2\,c_1^2
\sin^2{2\theta_3}, & for Cases II and
III.}}
With no further ado, we may set $s_1^2\simeq c_1^2\simeq 1/2$ in Eq.\econst\
from the Super-Kamiokande atnospheric neutrino
data~\ref\rfuk{Y. Fukuda
{\it et al.} [Super-K Collaboration], Phys. Rev. Lett. 81(1998)1562.}.

For Case I, Eq.\econst\ cannot provide a constraint on $\ee$ until an
experimental  lower
bound is set on the subdominant angle $|\theta_2|$. In the event that
 $\sin^2{2\theta_2}\simeq  0.1$,  its   largest 
CHOOZ-allowed~\ref\rCHOOZ{M. Apollonio {\it et al.,} Phys. Lett.
B466(1999)415.}~value, the constraint  $\ee<4.5\times 10^{-3}$
would be obtained. 
For Cases II or III, solar-neutrino data~\ref\rjb{{\it E.g.,} J.N. Bahcall,
M.C. Gonzalez-Garcia and C. Pe\~na, {\tt hep-ph/0212147.}}~determine 
$\sin^2{\theta_3}>0.68$ with 95\% confidence. Here the
stronger constraint $\ee< 1.7\times 10^{-3}$ results.

The non-orthogonality of $\nu_\mu$ and $\nu_\tau$ has another interesting
consequence. `Young' $\nu_\mu$'s (prior to the onset of oscillations) 
can act as if they were $\nu_\tau$'s
(thereby producing $\tau$'s via charged-current interactions)
with probability $\ee_\mu\ee_\tau$. Thus the unsuccessful NOMAD 
search~\ref\rnomad{P.Astier {\it et al.} [NOMAD Collaboration],
 Nucl. Phys. B611(2001)3.}\
for this process yields the inequalities:
\eqn\econstr{6.8\times 10^{-4}>
\cases{\ee^2\,c_2^4\sin^2{2\theta_1}, & for Case I;\cr  
 \ee^2\,c_3^4\sin^2{2\theta_1}, & for Cases II and
III.}}
For Case I, Eq.\econstr\ establishes the firm but weak upper limit
$\ee<2.6\times10^{-2}$.  For Cases II and III, the bound on $\ee$ set by
the NOMAD result is less stringent than our earlier result. 
\vfill\eject
To summarize, I have shown that current experimental data establishes the
bounds:
\eqn\esumm{\ee<
\cases{2.6\times 10^{-2}, & for Case I;\cr  
       1.7\times 10^{-3} & for Cases II and
III.}}
To illustrate the significance
of these results, note that the currently reported value~\rz\ 
for the
number of light neutrinos is $N_\nu=2.984\pm 0.008$. Its explanation 
in the present context (should a mere 2-sigma discrepency  need explanation)
  would indicate  $\ee= 4\pm 2\times 10^{-3}$. A similar remark probably
  applies to a discussion of the NuTeV anomaly. Thus, Case I 
is likely to provide  the only
feasible route to  effective  suppressive
mixing  in the minimal seesaw scheme.\footnote{${}^*$}{Other schemes of
  suppressive neutrino  mixing, including those with flavor-independent
suppression, can emerge from  more elaborate seesaw frameworks, such as one
  involving six heavy singlet meutrino states rather than three.}

\bigskip\bigskip
This work was supported in part by the National Science Foundation,
under grant NSF-PHY-0099529.

\listrefs
\bye